\documentclass{aastex}

\newcommand{\kms}{km~s$^{-1}$}

\slugcomment{(Version 2: Revision in the period analysis of the one suspected binary; the general results remain the same.) $-$ Scheduled to appear in Astrophys J, 734 (2011 June 10)}

\shorttitle{Radial Velocity Variations in PPNs}

\shortauthors{Hrivnak et al.}

\begin{document}

\title{Are Proto$-$Planetary Nebulae Shaped by a Binary? \\
Results of a Long-Term Radial Velocity Study\altaffilmark{1} }

\author{Bruce J. Hrivnak\altaffilmark{2}, Wenxian Lu\altaffilmark{2}, Katrina Loewe Wefel\altaffilmark{2,3}, David Bohlender\altaffilmark{4}, S. C. Morris\altaffilmark{4}, 
Andrew W. Woodsworth\altaffilmark{4},  and C. D. Scarfe\altaffilmark{5}}

\altaffiltext{1}{Based on observations made at the Dominion
Astrophysical Observatory, Herzberg Institute of Astrophysics,
National Research Council of Canada} 
\altaffiltext{2}{Department of Physics and Astronomy, Valparaiso University, 
Valparaiso, IN 46383; bruce.hrivnak@valpo.edu. wen.lu@valpo.edu}
\altaffiltext{3}{Present address: Parkland College, Champaign, IL 61821; kwefel@parkland.edu}
\altaffiltext{4}{Dominion Astrophysical
Observatory, Herzberg Institute of Astrophysics, National Research
Council of Canada, 5071 West Saanich Road, Victoria, BC V9E 2E7, Canada; David.Bohlender@nrc-cnrc.gc.ca} 
\altaffiltext{5}{Department of Physics and Astronomy, University of Victoria,
Victoria, BC V8W 3P6, Canada; scarfe@uvic.ca} 

\begin{abstract}

The shaping of the nebula is currently one of the outstanding unsolved problems in planetary nebula (PN) research.
Several mechanisms have been proposed, most of which require a binary companion.
However, direct evidence for a binary companion is lacking in most PNs.
We have addressed this problem by obtaining precise radial velocities of seven bright proto-planetary nebulae (PPNs),
objects in transition from the asymptotic giant branch to the PN phases of stellar evolution.
These have F-G spectral types and have the advantage over PNs of having more and sharper spectral lines, leading to better precision.
Our observations were made in two observing intervals, 1991-1995 and 2007-2010, and we have
included in our analysis some additional published and unpublished data.
Only one of the PPNs, IRAS 22272+5435, shows a long-term variation that might tentatively 
be attributed to a binary companion, with P $>$ 22 years, and from this, limiting binary 
parameters are calculated. 
Selection effects are also discussed.
These results set significant restrictions on the range of possible physical and orbital properties of any
binary companions: they have periods greater than 25 years or masses of brown dwarfs or super-Jupiters.  While not ruling out the binary hypothesis, it seems fair to say that these results do not support it.

\end{abstract}


\section{INTRODUCTION}

\subsection{\bf Background to the Question}
\label{intro}

Arguably the most controversial area in planetary nebula (PN) research at present is
the determination of the mechanism for shaping the nebula.  
This has been brought to the  fore by the visually stunning, high-resolution 
{\it Hubble Space Telescope} ({\it HST}) images.  
PNs generally possess an elliptical or bipolar structure, often with 
additional point symmetric features \citep[ansae, jets; see][]{bal02}.
In contrast, their precursors, asymptotic giant branch (AGB) stars, have been considered to be
basically spherical \citep{olof99,neri98}, 
although some recent resolved molecular-line observations show that a significant
fraction show some axial symmetry \citep{castro07}.
So the question can be posed as to how a mass-losing AGB star, which is basically spherical,
can evolve into the variety of PN shapes.

The presently emerging consensus is that the axial and bipolar asymmetry is 
caused directly or indirectly by the presence of a binary companion 
to the central star.  
A binary companion could influence the density structure in at least the following three ways.  

\vspace{-0.1in}
\begin{enumerate}
\item Most directly, a companion could gravitationally focus 
the mass loss into the orbital plane, forming an equatorial density enhancement 
and perhaps a torus. 
This density enhancement or torus would then collimate the fast wind producing a bipolar
outflow \citep{liv88}.  A variation on this would be the
formation of an accretion disk around the binary companion; this could
collimate a fast wind and carve out the lobes
and also  lead to point-symmetric ejecta \citep{sokrap00}.
\vspace{-0.1in}%
\item The mass could be preferentially lost in the equatorial plane of the PN 
during the AGB phase due to a rotationally-induced oblate shape, 
producing a collimating torus \citep{garseg99}.
However, given the low rotational velocity of an intermediate-mass star on the main sequence,
it would probably require the presence of a close companion to spin up the resulting AGB star
by the transfer of orbital angular momentum.
This binary-induced mass loss could occur during a common envelope phase \citep{nor06}.
\vspace{-0.1in}%
\item The central star could possess a magnetic field that
collimates the outflow into bipolar lobes \citep{garseg99,garseg05}. 
Recent work indicates that something like a binary interaction 
would most probably be needed to sustain such a magnetic field 
through the transfer of angular momentum \citep{nor07}. 
\end{enumerate}

The first mechanism is generally favored to produce the density enhancement
or collimating torus, although it can be seen that a binary companion
would be important in all three mechanisms. 
Population synthesis studies suggest that binary interactions 
can produce the correct number of Galactic PNs \citep{moe06}.
Thus it is increasingly common to hear it stated that elliptical or bipolar PNs 
are due to the effect of a binary companion, or even that the
presence of an axially-symmetric or bipolar structure implies a binary. 
However, this hypothesis has not been adequately supported observationally. 
An extensive review of the question of what shapes PNs, 
with a thorough investigation of the evidence for the binary hypothesis, 
has recently been presented by \citet{demar09}; 
the need for direct observational tests of this hypothesis is stressed
\footnote{To address this outstanding problem in the study of PNs, an informal 
international collaboration was recently formed with the goal of settling the 
question of the binarity in PNs and its effect on the shaping of the
nebula (PlaN-B; coordinator O. De Marco; http://www.wiyn.org/plan-b/).}.

\subsection{Why Search for Binary PPNs?}

Proto-planetary nebulae (PPNs) are the objects in transition between the 
AGB and PN phases in the evolution of intermediate- and low-mass stars.  
During the AGB phase, such stars are surrounded by an expanding circumstellar
envelope CSE of mass being lost at an increasing rate.  In the
PPN phase, the high rate mass loss has ended and the star is
surrounded by a detached, expanding envelope \citep{kwo93}.
PPNs are a subset of the larger class of post-AGB stars, 
which also includes RV Tau and R CrB variables \citep{vanwin03}.
PPNs can be distinguished by having more massive circumstellar nebulae and
in many cases showing clear abundance patterns from AGB nucleosynthesis.
As such, PPNs appear to be the most likely objects to evolve into PNs.
In the Discussion section, we will distinguish in more 
detail PPNs from post-AGB stars in general and compare our results with those 
found for other post-AGB stars.  

PPNs display a basic axial symmetry, often showing bipolar lobes and occasionally point symmetry.  
This has been particularly seen in high$-$resolution {\it HST} images \citep{uet00,su01,sah07,sio08}.  
Some also display an obscured equatorial region.  
Thus one sees in PPNs the same basic structures as in PNs, but at an earlier stage in the nebula, 
a stage closer to the beginning of the shaping process.
This commends the study of PPNs to investigate the shaping mechanism(s).

Binarity can manifest itself in several ways: a visible companion, photometric light 
variations, composite spectrum, and radial velocity variations.
A survey of the results of these methods has been presented
earlier \citep{hri09a} with null results;
no evidence of binary companions to PPNs was found.

In this paper, we discuss the observational evidence of binarity in PPNs 
based on long-term radial velocity studies 
of seven bright PPNs. 
The results of this study and their implications for the binary nature of PPNs are then
discussed, 
and conclusions drawn and discussed on whether or not they provide evidence to support the
binary hypothesis. 

\section{OBSERVATIONAL SAMPLE AND DATA SETS}
\label{sample}

Our sample consists of the seven brightest PPNs observable from mid-latitudes 
in the northern hemisphere.  They range in visual brightness from 7.1 to 10.4 mag
and are listed in Table~\ref{ppn_prop}.
We examine later the question of what biases might result from observing the
brightest PPNs.
For uniformity, we will refer to all of them by their {\it IRAS}
identification numbers. 
All show the double-peaked spectral energy distribution characteristic 
of PPNs \citep[e.g.,][]{hri89}, and all have F$-$G spectral types.
All are known to vary in light and all but one have variable star designations.
Detailed light curve studies of all seven of these have been presented
elsewhere \citep{hri10,ark10,ark06,fer83}.

\placetable{ppn_prop}

The main data sets used in this study are the radial velocity observations we carried out
at the Dominion Astrophysical Observatory (DAO), initially from 1991-1995 and 
then more recently from 2007-2010.  
The 1991-1995 data were obtained with the radial-velocity spectrometer \citep[RVS,][]{fle82}, 
while those from 2007-2010 were derived from CCD spectra 
by cross-correlation with several IAU standards. However, the velocities of 
those standards were taken to be those derived at the DAO photographically 
\citep{sca90,sca10} and with the RVS. The zero-point of the much more 
numerous RVS observations of standard stars has been adjusted to match that of 
the photographic data for this purpose, this ensuring that the CCD and RVS 
data are on the same system.
With F$-$G spectral types, these objects have numerous sharp absorption lines, 
which result in an observational precision of $\sim$0.7 km~s$^{-1}$. 
The individual radial velocity measurements will be published elsewhere in the context of our
detailed pulsational studies of the individual objects.
We find that they all vary in velocity and light due to pulsation, with periods 
ranging from 35 to 130 days \citep[see also \cite{ark00, ark10}]{hri09b,hri10}.
Substantial radial velocity data sets have also been published by other investigators for three of the objects, 
IRAS 07134+1005 \citep{leb96,barth00}, 17436+5003 \citep{bur80}, 
and 22272+5435 \citep{zacs09}, as part of their study of pulsation in these objects.
These published data sets and a few individual published and unpublished velocity observations 
have also been incorporated into this study.  

\section{RESULTS OF THE BINARY INVESTIGATION}
\label{results}

\subsection{Investigating the Sample}

\begin{figure}\figurenum{1a}\epsscale{0.80} 
\rotatebox{180}{\plotone{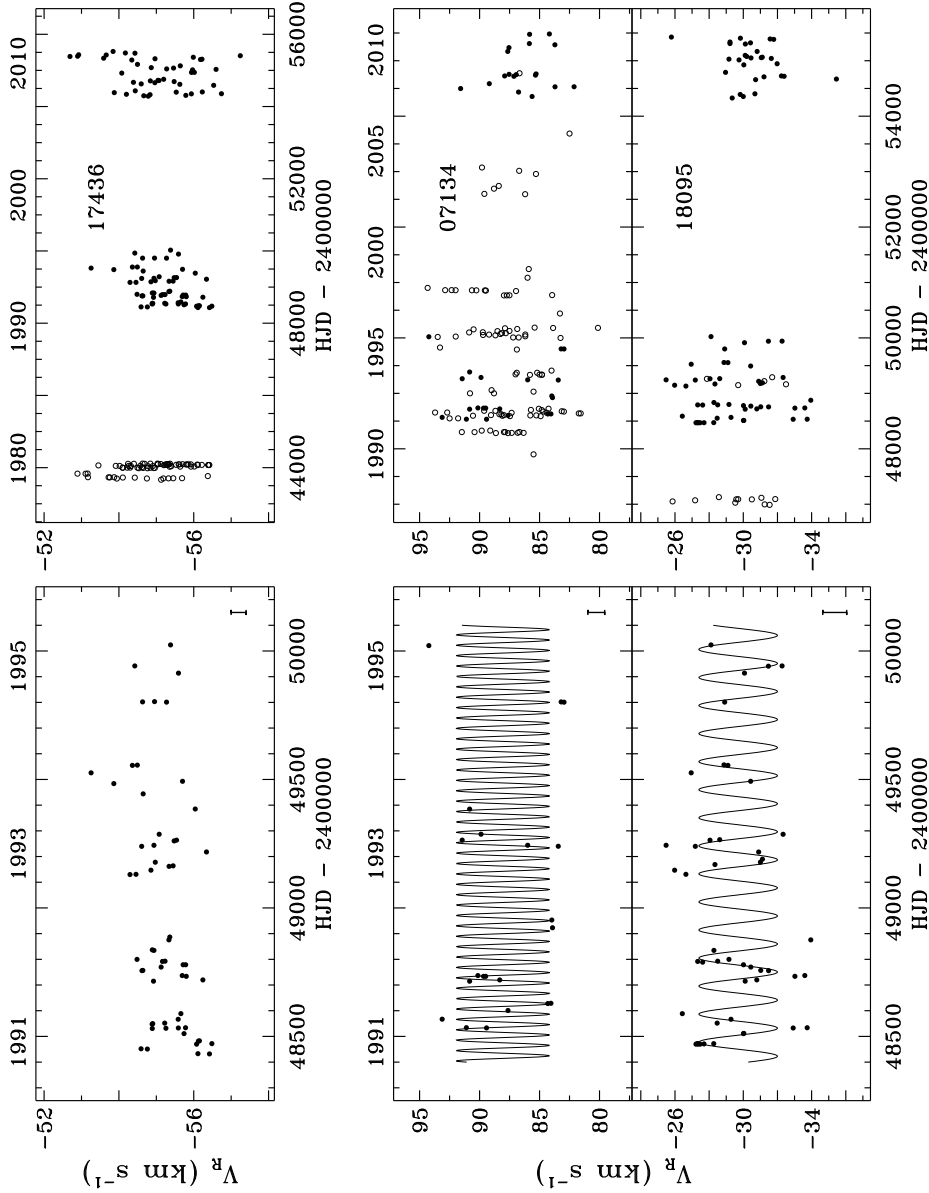}}
\caption{(Left) Radial velocity curves from our 1991$-$1995 observations, showing sine-curve fits to the periodic pulsational variations.  (Right) Radial velocity curves showing the long-term velocity variations, including both our observations (filled circles) and those of others (open circles).
Sample average error bars ($\pm$1 $\sigma$) are shown for each PPN in the lower right-hand corner of the left panels.
\label{vc_all1}}
\epsscale{1.0}
\end{figure}

\begin{figure}\figurenum{1b}\epsscale{0.9} 
\rotatebox{180}{\plotone{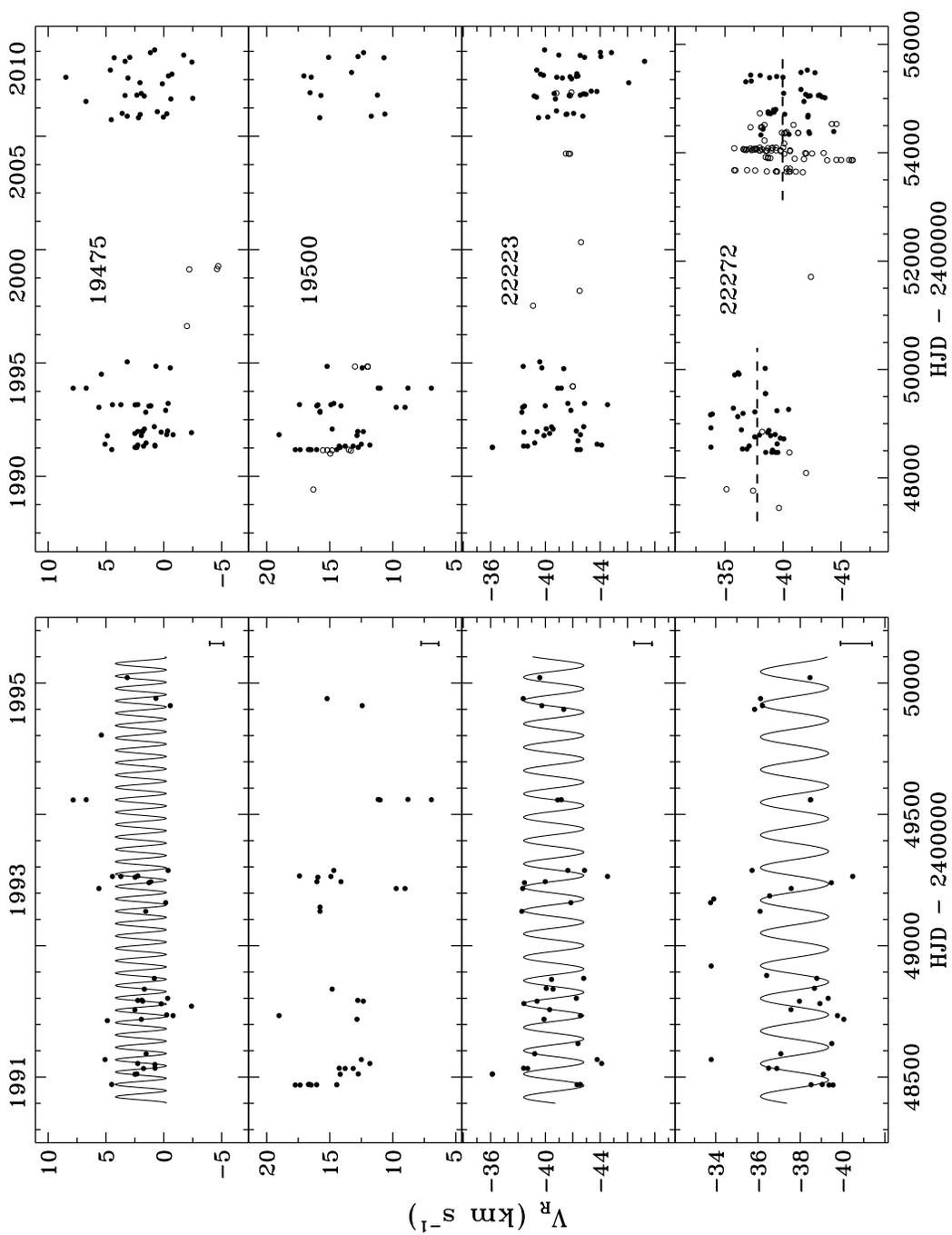}}
\caption{Radial velocity curves, with symbols similar to Figure 1a.
The dashed lines in the right-hand panel for IRAS 22272+545 represent the average velocities in the 
two observing intervals.  
\label{vc_all2}}
\epsscale{1.0}
\end{figure}

In Figures~\ref{vc_all1} and \ref{vc_all2} are plotted the radial velocity data for these seven objects.  In the left-hand panels are shown our observations from 1991-1995 
on a scale that allows one to get a better sense of the velocity variations, 
and in some cases the clear cyclical nature of the pulsations is indicated by sine-curve fits.
In the right-hand panels are shown the data over the entire range of observations, 
including those by others.
The combined data sets for each of the objects show a range of velocities of 
10 to 14 km~s$^{-1}$.

\placefigure{vc_all1}

\placefigure{vc_all2}

Comparing our 1991-1995 data with 2007-2010 data for the seven objects 
(summarized in Table~\ref{obs_rv}),
we find only one object, IRAS 22272+5435, for which the average values differ by more 
that 1.5 \kms (or 2$\sigma$).
For the others, the difference between the two data sets is $\le$~2$\sigma$ and 
it does not appear that the average values differ significantly between the two epochs of observation.
Nor is there evidence for systematic change when we include the other data sets.
A formal period analysis of all of the data was carried out for each object using the
Period04 \citep{lenz05} program.
Beyond the pulsation mentioned earlier, there is a suggestion of a long-term periodicity only  in 
IRAS 22272+5435, which will be discussed in Sec.~\ref{22272}.     

\placetable{obs_rv}

To investigate even longer-term variations, we compared the average radial velocity 
from these optical, photospheric spectra with the radial velocity determined from the 
circumstellar CO or OH emission.  
The CO and OH represent the emission from the circumstellar envelopes, which
have been expanding over several thousand years, since the late stages of the AGB phase.
Thus their velocity centers should be the same as that of the average PPN velocities,
which should be that of the optical PPN velocity if there is no binary motion 
and that of the barycenters if there is binary motion.
These results are also listed in Table~\ref{obs_rv}.
We see that these molecular-line velocities, when transformed to the heliocentric system, 
are all, with one exception, close to the velocities based on the visible spectra,
differing by 0 to $-$2 km~s$^{-1}$.
The lone exception is IRAS 18095+2704, which differs by +5 km~s$^{-1}$.
This is the only target detected in OH \citep{eder88}, and inspection of the spectrum shows that the two separate
maser components are relatively weak compared to most of the others in the study and that they show structure.
\citet{buj92} cite a tentative CO detection at a velocity that would reduce the difference to +2 km~s$^{-1}$.
Based on these data, we do not see any evidence from the molecular-line observations to indicate a significant difference
between the photospheric and the circumstellar envelope velocities, although there appears to be some inconsistency 
with the OH velocity of IRAS 18095+2704.

\subsection{Evidence for a Possible Binary Companion to IRAS 22272+5435} 
\label{22272}

A comparison of the earlier (1989-1995) and later (2005-2010) observations of 
IRAS 22272+5435, including the data from \citet{zacs09}, shows a bimodal distribution of velocities.  There exists a significant difference of $-$2.2 km~s$^{-1}$ in the average 
velocities of the data sets from the two epochs, about four times the sum of the uncertainties in the values. 
We interpret this velocity difference as most probably due to the  motion of the PPN around the barycenter of a binary star system.
We note that IRAS 22272+5435 is the coolest and reddest star in the sample 
and presumably has the lowest surface gravity, as indicated by it spectral type and 
the results of a model atmosphere analysis \citep{red02}.  
Hence it is likely to be the least stable on a time scale of many years.
While we have considered this possibility, we still think that the binary hypothesis
is more likely to explain the long-term velocity variation.

A formal period study was carried out of all of these data.  
In addition to the pulsation period of 128.3 $\pm$ 0.1 days, 
no reliable long-term period could be determined. 
This is not surprising, give the distribution of the velocities in the two 
observing intervals around two nearly constant average velocities,
$-$37.8 km~s$^{-1}$ for the 1989$-$1995 observations and 
$-$40.0 km~s$^{-1}$ for the 2005$-$2010 observations,
with no transition between them.

This does not mean, however, that we have no idea of period of the suspected binary.
Given the observing intervals, it would not be possible for a binary to go through a 
complete cycle of variability in a time less than the total observing interval of 22 years.
Thus we have a minimum value for the period of the suggested binary orbit.
Continued monitoring will help to further constrain this by revealing a period or increasing 
the minimum value.

Thus on the basis of the change in the radial velocities of IRAS 22272+5435, 
we make a tentative, but we think probable, identification of a binary companion.
If we make the assumption that the long-term velocity difference is due to binary motion,
we can then carry out a radial velocity solution and
determine limiting binary parameters.
For this we use the minimum values of 22 years and 1.3 km~s$^{-1}$
for the values of the period and observed velocity semi-amplitude,
the latter being half of the difference in the average velocity of the two intervals
when the 128.4 day period is removed from the data.
Assuming a circular orbit, M$_{\rm PPN}$=0.62 M$_\sun$
(which appears to be a typical post-AGB core mass based on models for a star 
with an initial main sequence mass of 2-3 M$_{\sun}$ \citep{blo95,vas94}), 
and a range of inclinations, one finds from the 
mass function a range of possible masses for the secondary: M$_2$ =  
0.10 M$_\sun$ ({\it i}=90$\arcdeg$), 0.12 M$_\sun$ ({\it i}=60$\arcdeg$),
0.22 M$_\sun$ ({\it i}=30$\arcdeg$), and 0.51 M$_\sun$ ({\it i}=15$\arcdeg$).
From a mid-infrared imaging and modeling study, \citet{uet01} determined 
an inclination of the torus of {\it i}=25$\arcdeg$$\pm$3$\arcdeg$.
Assuming that the binary orbit is coplanar with the equatorial density enhancement, 
this leads to a companion mass of 0.27$\pm$0.04 M$_\sun$.
This is a very reasonable value for a secondary companion, and the assumption that it is 
a main-sequence star would place it at a spectral class of $\sim$M4 V (Cox 2000).  
For a circular orbit, this leads to a separation of 8.0 AU = 1700 R$_\sun$,
a value not only well outside the present radius of the PPN of R$\sim$100 R$_\sun$,
but one that would have been well outside it when the star was at the tip of the AGB,
with R$\le$500 R$_\sun$.
We emphasize that these are only preliminary, minimum values, based on the hypothesis that this 
long-term velocity difference is due to binary motion.  
For example, if this is a binary with a period of 34 years and the same semi-amplitude as above,  
the calculated value for the mass would increase to 0.32 M$_\sun$, and if the velocity 
semi-amplitude were 1.8 km~s$^{-1}$, the mass would increase to 0.50 M$_\sun$.
Calculations have shown that a binary with such parameters can form a very narrow-waist
bipolar PN \citep{sokrap00}, although in this case the visible {\it HST} image shows the nebula to
not be this extreme in shape.
These nebulae are all surrounded by a larger halo, representing the earlier AGB mass loss.
\citet{hug10} calculate the effects of different parameters ({\it M$_2$, i, a}) on the shaping of
an initially spherical AGB wind to produce the observed halo in a PPN.  
They run a model for a binary with parameters similar to the preliminary values 
that we found for IRAS 22272+5435, 
for which they predict an approximately round halo.  This is consistent with that 
observed for this PPNe \citep[classify the halo as centrosymmetric with arc-like features]{sah07}. 
(Given an estimated distance of 1.9 kpc and a luminosity ratio
of $\approx$3$\times$10$^5$ based on the M4~V spectral classification, there is no present hope 
of observing the companion directly.)

\subsection{Selection Effects and Limits on Binarity from the Null Results}

For the other six PPNs, no systematic change in velocity has occurred 
between the 1991$-$1995 and the 2007$-$2010 observations, nor is any seen 
when we include additional published velocities. 
Period analyses show only the pulsational velocity variation.
However, one needs to consider selection effects.
Since we observed the
brightest PPNs, they are not highly obscured.
Thus, if they have a bipolar structure with an enhanced equatorial density region, 
they are probably biased towards having their equatorial planes close to 
the plane of the sky, thereby reducing their obscuration.
This would 
reduce the observed orbital velocity, 
given the reasonable assumption that the binary orbit is 
in the equatorial plane.

\begin{figure}\figurenum{2}\epsscale{0.75}
\rotatebox{90}{\plotone{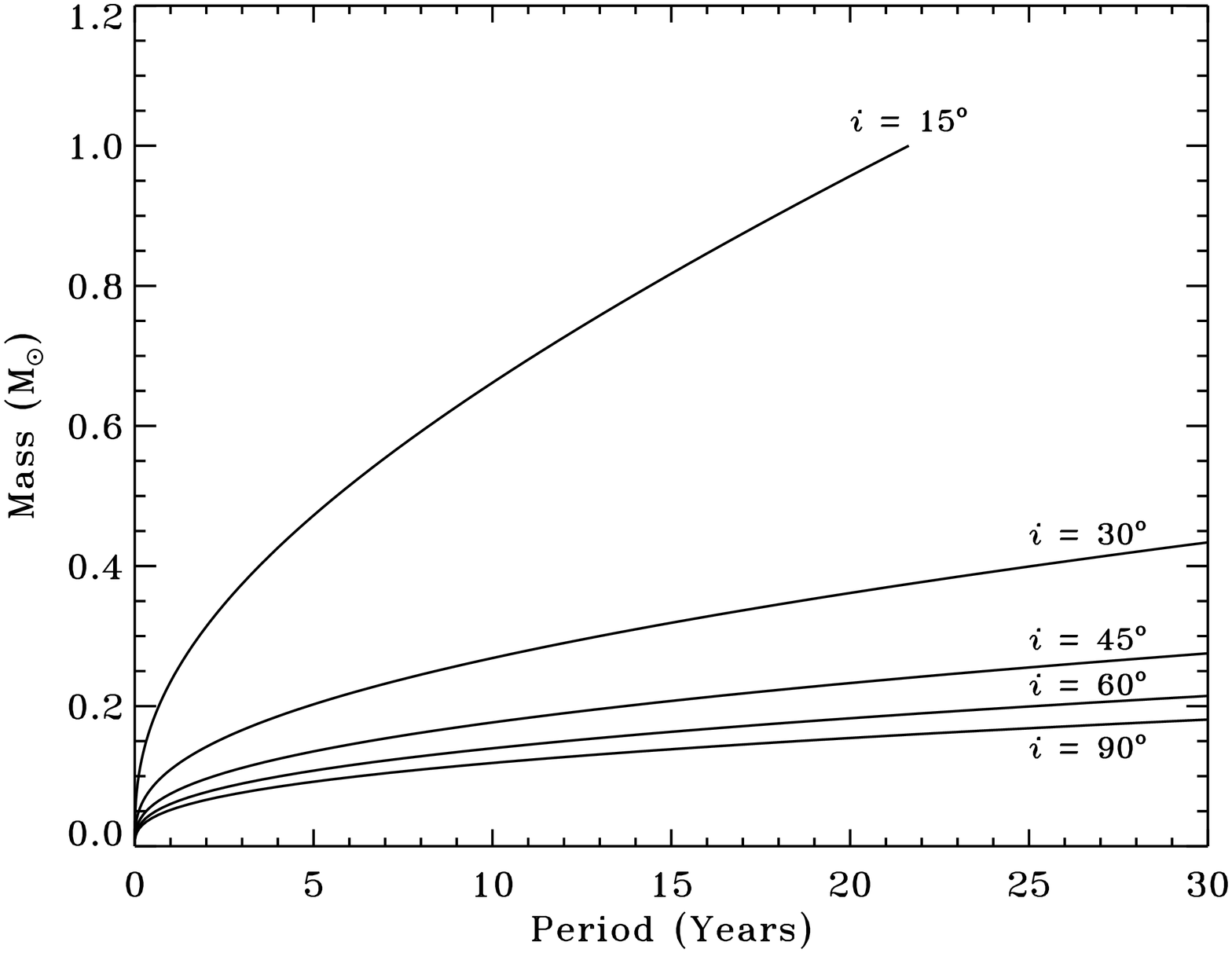}}
\caption{Plot showing limits on M$_2$ and P for various inclinations (marked on the curves), assuming M$_{\rm PPN}$=0.62 M$_\sun$ and K$\le$2.0 km~s$^{-1}$, based on the null results for binary detection for the other six PPNs.  To remain undetected, any binaries present in the sample must have companion masses lower than or periods longer than the values of the curves at the various inclinations (i.e., they must lie below the curves).
\label{vc_limits}}\epsscale{1.0}
\end{figure}

In spite of the uncertainty in the inclination of the equator 
(which mid-infrared images and spatial-kinematical observations can help solve), 
these null results still set useful limits on the binary parameter space. 
These are shown graphically in Figure~\ref{vc_limits},
assuming that M$_{\rm PPN}$=0.62 M$_\sun$, circular orbits,  
and a conservative detection limit of K=2.0 km~s$^{-1}$.
This shows that to remain undetected, 
a companion of 0.40 M$_\sun$ must have an orbital period
of P$>$3.5 yr if {\it i}$\ge$15$\arcdeg$ and P$>$24.5 yr if {\it i}$\ge$30$\arcdeg$;
an undetected  companion of 0.25 M$_\sun$ must have an orbital period
of P$>$1.1 yr if {\it i}$\ge$15$\arcdeg$, P$>$8 yr if {\it i}$\ge$30$\arcdeg$, and
P$>$23 yr if {\it i}$\ge$45$\arcdeg$.
Another way to look at it is that M$_2$$>$0.65 M$_\sun$ is excluded for P$<$10 yr 
at {\it i}$\ge$15$\arcdeg$ and M$_2$$>$0.30 M$_\sun$ is excluded for P$<$13 yr 
at {\it i}$\ge$30$\arcdeg$.
Lower-mass or longer-period companions than these would escape detection.

\placefigure{vc_limits}

We do know something about the orientation of the bipolar
structure in these six PPNs based on {\it HST} images and on 2-D modeling
in several cases with resolved mid-infrared images.  {\it HST} images of IRAS
18095+2704 and 19475+3119 suggest that they are viewed at some
intermediate angles \citep{uet00,sah07}.  Models based in mid-IR images 
and, for IRAS 07134+1005, spatially-resolved molecular-line spectroscopy,
have determined the approximate inclinations of the polar axis with respect
to the plane of the sky for two of the objects.
For IRAS 07134+1005, {\it i} $\sim$ 80$\arcdeg$ \citep{mei04}, 
and for IRAS 17436+5003, {\it i} $\sim$ 90$\arcdeg$ \citep{mei02} or {\it i} $\sim$ 10$\arcdeg$
\citep{gle03}, although for the latter of these objects one sees very disparate results.
For IRAS 22223+4327, a comparison of mid-infrared and {\it HST} visible
images implies an inclination that is closer to edge-on (90$\arcdeg$) than pole-on \citep{clu06}.
For IRAS 19500$-$1709, the inclination is more uncertain 
\citep[Volk, personal communication 2008]{gle01}.
Therefore at least some of these appear to be inclined out of the plane of the sky
with {\it i} $>$ 30$\arcdeg$,  and there does not appear to be an important bias of the sample to the plane of the sky.
In this case the projection effects are less severe and the constraints 
on the mass or period of any undetected binary companion are more significant.
Thus it appears that our non-detection of binarity in all six of these PPNs cannot be
explained away as simply due to a low binary inclination to the plane of the sky.

\section{DISCUSSION AND CONCLUSIONS}
\label{discuss}

This radial velocity study of seven PPNs shows 
only one has velocity variations, in addition to pulsation,
that can tentatively be attributed to a binary companion.  
The presence of pulsation does complicate the search for a binary companion and 
makes it more difficult, but it does not invalidate the null result for these other six.
As a counter example, we found for the related post-AGB object 89 Her that we could
detect a binary companion (K$=$3.3 km~s$^{-1}$, P$=$290 d) even
with a pulsating central star (K$=$1.6 km~s$^{-1}$, P$=$66 d; in preparation).
Thus such binaries could be detected but were not.
We used these null results to set limits on the properties of any 
undetected companions.  
The one tentative binary has a long period ($>$22 years) 
and probably a normal stellar mass companion (M$_2$$>$$\approx$0.27 M$_{\sun}$).

One might initially be surprised by this low binary fraction in light of the 
discovery of a large number of post$-$AGB binaries by Van Winckel and collaborators
\citep{vanwin03,vanwin07,vanwin06}. 
They find thus far that 27 out of a sample of 51 post-AGB stars are spectroscopic binaries.
However, these 51 post-AGB are not an unbiased sample but were chosen because they possess 
several of the observed characteristics of previously known post-AGB binaries \citep{der06}.
These post-AGB binaries are a distinctly 
different set of objects than the PPNs.  They show a broad infrared excess (broad SED),
indicating both hot and cool dust, and have abundance anomalies
thought to be due to chemical fractionation of refractory elements
onto dust, with re-accretion of non-refractory elements by the
star \citep{vanwin03}. These properties are attributed to the presence of a
circumbinary disk. Most and perhaps all of these objects are
binaries, with P $\approx$ 100$-$2600 days and e = 0.0$-$0.6.
We would have detected such binaries but did not.
In these post-AGB binaries, it is the binary that is thought to be responsible for forming and 
stabilizing the disk \citep{vanwin06}. 
The orbital periods of the shorter of these are of such that the systems 
would not accommodate within them a large AGB star.
Thus it appears that it is their binary nature that
leads to their special properties \citep{wae04} and brings them to our
attention due to their infrared excesses.   
 PPNs do not share
these properties, but rather display a clearly double-peaked SED
with a much larger infrared excess, indicating a detached shell and much larger
mass loss.
The PPNs, at least the carbon-rich ones, have abundances in 
agreement with AGB nucleosynthesis, and none have the
abundance anomalies seen in the post-AGB binaries.
Also, the PPNs show a visible nebula which most of the post-AGB 
binaries do not (the Red Rectangle is an exception ).
Thus, in contrast to \citet{demar09}, we conclude that the binary post-AGB objects 
in general represent a class of 
objects that are unlikely to evolve into PNs and therefore 
do not bear directly on the question of the shaping of the nebulae.
PPNs, on the other hand, give every indication of being the immediate precursors 
of PNs. 

We can make the comparison instead to the binary 
central stars of PNs.  
Photometric searches indicate that 10$-$20 \% of all PNs have a close (P$<$8 d) companion
\citep{misz09a,demar09}.
With their short periods, it seems likely that the binary PNs formed through 
common envelope evolution in which the two stars did not merge.  
Might the PPNs be binaries, but presently in the common envelope
stage?  Since the common envelope stage is calculated to be very short, on the
order of the pre-common envelope orbital period \citep{ric08}, 
this cannot be the case, for it would be far too improbable to find 
six of our seven in this very short-lived phase.
We cannot make a comparison with the fraction of PNs with a period in the range of 
0.1 to 30 years, since this is observationally unknown.
Radial velocity studies of PNs with a resolution of 3 km~s$^{-1}$ have  
been initiated \citep{dem06}.  However, these are complicated by the broad 
lines in the central stars and their variable winds, and no definitive results have been obtained.  

What do we know about the shapes of the binary PNs and how do they compare with the
shapes of these PPNs?
Based on a sample of 30 of these binary PNs with good images, it has been determined 
that $\sim$30$\%$ have nebulae with clear bipolar morphologies, and it is suggested that
this number might be as high as $\sim$60$\%$ if inclination effects and other factors are
included \citep{misz09b}.  
This result is highly suggestive that a common envelope evolution without merger
will commonly produce a bipolar nebula.  
However, this does not imply the inverse, that bipolar nebulae have a binary central star.
All of our seven PPNs have a bright central star and would be classified as SOLE in the classification
scheme of \citet{uet00} and \citet{sio08}.
In the more detailed classification scheme of \citet{sah07}, six of the seven are classified:
four as elongated, one as bipolar, and one as multipolar (see Table~\ref{ppn_prop}).
This suggests they each have an axis of symmetry that might arise from a equatorial 
density enhancement, and in several of the cases this enhancement is seen in the mid-infrared images.

One is still left with the question of why only one of our seven PPNs shows 
evidence of being a binary, given that the binary fraction of stars is so much higher.
The careful study of a sample of 164 solar-type (F7$-$G9 IV-V, V, VI) stars
by \citet{duq91} finds $\sim$50 \% to be binaries.
But this included visual binaries and common proper motion pairs, 
and resulted in a mean period of 180 years.
If we restrict our comparison to the spectroscopic binaries, the fraction drops to $\sim$25 \% 
(with orbits) or $\sim$33 \% (including those detected to vary in velocity as
binaries but without determined orbits).  These results are based on 
high-precision velocities ($\sigma$ $<$ 0.3 km s$^{-1}$) over an observing
range of up to 13 years (average 8.6 years) and are a better comparison with
our radial velocity sample.  
Given their higher precision and the absence of the pulsational variations 
which complicate the study of our stars, our tentative detection of one in seven (14 \%) to be binaries 
does not appear to be anomalously low.

Might these PPNs be binaries but with periods longer than 25 yr?  These
might not be detected in this radial velocity study, 
but they could still affect the shaping of the nebula, although their effect would 
be reduced with increasing orbital period and separation. 
Might they be binaries but with low-mass ($<$0.25 M$_\sun$) companions?
Our above limits on binarity do not exclude such companions. 
If the companion is a brown dwarf or a super-Jupiter planet, then it would escape detection in
our program.  
While these can have a significant effect on the mass loss in certain cases and produce 
elliptical nebulae, it is estimated that planets will significantly affect mass loss in only 
4--10$\%$ of AGB stars \citep{liv02}

The results of this present radial velocity study provide the first direct test
of the binary hypothesis in shaping PPNs, the direct precursors of PNs.
While they do not rule it out,
it seems fair to say that they do not support the binary hypothesis.
Although this study has not answered the question of whether the shaping
of PPNs and PNs is ultimately due to a binary companion, it has set 
significant constraints on the properties of a binary companion during the
PPN phase.
The lack of detection of a companion in six of the cases probably implies that 
any such companion either has a period that is very long ($>$25 yr) or  
a mass that is very low ($<$0.25 M$_\sun$).  
The effects of these on the mass loss and its shaping are obviously less than 
in the case of a higher mass, shorter period companion.
These constraints can help guide future attempts to model the formation
of the circumstellar density asymmetries with a binary companion.
And of course they might not be binaries, and the asymmetric
mass loss would then be due to something else.  
We know that they were
pulsating during the previous AGB phase, and perhaps pulsation coupled 
with some other mechanism such as cool star spots \citep{sok00} is the mechanism 
responsible for the shaping.
These results also do not appear to support the hypothesis that the intensive mass
loss at the end of the AGB (the ``superwind'') is driven by a binary companion 
\citep{demar09}.  
Since in these seven PPNs it is apparent that the envelopes are detached and 
the shaping of the nebula
has started, these results might suggest two ways to form the
shapes seen in PNs: (a) through common envelope evolution, as
evidenced by the close binary nuclei of some PNs, or (b) through a
non-common envelope process, which is occurring in these PPNs.
This latter process might involve a distant and/or low-mass
companion or be due to a single, pulsating central star.

This radial velocity study is continuing so that we can extend the
temporal baseline in the search for evidence of even longer period binaries 
and seek to confirm the one tentative case.
We have also begun a radial velocity study of several edge-on
bipolar PPNs, in which we make the reasonable assumption that the
binary orbit would be in the plane of the equatorial density enhancement.  
In such a case, we would see the full orbital velocity variations without suffering
from an inclination effect.  
While in these cases the star is completely obscured from view in
visible light, it is seen in the near-infrared and thus amenable to 
near-infrared spectroscopy.
These PPNs have bipolar lobes and an obscuring dust lane, implying very 
strong shaping of the outflow.
Rotation will also be investigated by comparing these edge-on cases
with ones that are more nearly pole-on; 
since the pole-on ones are expected to appear as slow rotators, 
this comparison can give evidence in the edge-on cases 
of possible rotational spin-up or merger by a companion.

\acknowledgments  We thank R. McClure and C. Waelkens for sharing their
unpublished observations with us, D. Crabtree for making some observations for us,
and D. Westpfahl for making some initial observations which showed the
variability of one of these objects.
We thank T. Hillwig for helpful comments on the manuscript and the referee for
critical comments that helped improve the the focus and avoid over-interpreting the
velocity data for IRAS 22272+5435.
BJH acknowledges the support of a University Research Professorship and a 
sabbatical leave from Valparaiso University and the hospitality of the 
Dominion Astrophysical Observatory during the initial stages of this research project.
BJH also acknowledges support from the National Science Foundation
(AST 9018032, 0407087, 1009974) and the NASA Astrophysics Data Program.

\clearpage

\begin{deluxetable}{rrrcrcc}
\tablecaption{List of Program Objects\label{ppn_prop}}
\tabletypesize{\footnotesize} \tablewidth{0pt} \tablehead{
\colhead{IRAS ID}&\colhead{HD}&\colhead{Var. Star}&\colhead{Other Names}
&\colhead{V(mag)}&\colhead{Sp.T.}&\colhead{Morph. Class.\tablenotemark{a}}} \startdata
07134$+$1005 &   56126 & CY CMi  & LS VI +10 15&  8.2 & F5 I & Ec*(0.41),h(e)\\
17436$+$5003 &  161796 & V841 Her & \nodata &  7.1 & G3 Ib & Ec*(0.41) \\
18095$+$2704 & \nodata & V887 Her & \nodata  & 10.4 & F3 Ib & Bc,h \\
19475$+$3112 & 331319 & \nodata & LS II +31 9   &  9.4 & F3 I & Mc*(0.43),ps(m,s),h \\
19500$-$1709 &  187885 &  V5112 Sgr & \nodata   &  8.7 & F3 I & \nodata\\
22223$+$4327 & \nodata & V448~Lac & DO 41288 & 9.7 & G0 Ia & Ec*(0.43),h(a) \\
22272$+$5435 &  235858 & V354~Lac & \nodata & 9.0 & G5 Ia & Ec*(0.55),h(a) \\
\enddata
\tablenotetext{a}{According to the classification system of \citet{sah07}.  The main classifications are E - elongated, B - bipolar, M - multipolar; {\it c} indicates closed lobes, * that the star is visible, {\it ps} the presence of point symmetry, and 
{\it h} the presence of a halo.  For more details, see their paper.}
\end{deluxetable}

\clearpage

\begin{deluxetable}{rrrrrrrrrrrl}
\rotate \tablenum{2} \tablecolumns{13}
\tabletypesize{\footnotesize}\tablewidth{0pt} 
\tablecaption{Summary of Radial Velocity Observations
\label{obs_rv}} 
\tablehead{
\colhead{IRAS ID} & \multicolumn{3}{c}{Number of Observations} &\colhead{}&\multicolumn{3}{c}{$<$V$_R$$>$
(km~s$^{-1}$)\tablenotemark{a}}
&\colhead{$\Delta$V$_R$\tablenotemark{b}}&\colhead{V$_{LSR}$(CO,OH)\tablenotemark{c}}&\colhead{V$_R$(CO,OH)\tablenotemark{d}} & \colhead{Comments} \\
\cline{2-4}\cline{6-8}
\colhead{}&\colhead{1991-95\tablenotemark{e}}&\colhead{2007-10\tablenotemark{e}}&\colhead{Total\tablenotemark{f}}&\colhead{}
&\colhead{1991-95\tablenotemark{e}}&\colhead{2007-10\tablenotemark{e}}&\colhead{Total\tablenotemark{f}}&\colhead{(km~s$^{-1}$)}
&\colhead{(km~s$^{-1}$)}&\colhead{(km~s$^{-1}$)}& \colhead{} }
\startdata
07134$+$1005 & 21 & 18 & 141 &&  88.0(0.8) &  86.3(0.5) &  87.5(0.3) &  14 & 73& 88 & \nodata \\
17436$+$5003 & 59 & 45 & 178 &&  $-$53.1(0.2) &  $-$52.6(0.4) &   $-$52.6(0.2) & 11 & $-$35 & $-$54 & longer P? \\
18095$+$2704 & 47 & 29 &   77  &&  $-$29.4(0.3) & $-$30.5(0.3) &  $-$29.8(0.2) & 10 & $-$5 & $-$25 & \nodata \\
19475$+$3112 & 38 & 29 &   71  &&  2.1(0.4) &  1.9(0.5) &  1.7(0.3) &  13 & 18 & 0   & \nodata \\
19500$-$1709 & 35 &  13 &  58 &&  14.5(0.5) &  13.8(0.7) & 13.9(0.3) &  12 & 25& 13  & \nodata  \\
22223$+$4327 & 34 & 36 &   81 &&  $-$40.5(0.4) &  $-$41.9(0.3) &  $-$41.3(0.2) & 11 & $-$30 & $-$42 &\nodata \\
22272$+$5435 & 34 & 36 & 155 &&  $-$37.6(0.3) &  $-$40.6(0.3) & $-$39.4(0.2) & 12 & $-$28 & $-$40 &  longer P \\
\enddata
\tablenotetext{a}{The values in parentheses represent the uncertainties in the mean values.}
\tablenotetext{b}{Range in radial velocities, based on the total data set.}
\tablenotetext{c}{CO or OH molecular-line velocities in the local standard of rest ({\it LSR}) system. References: CO -- \citet{lik91,omont93,buj92,hri05}; OH -- \citet{eder88}. }
\tablenotetext{d}{V(CO,OH) transformed from the {\it LSR} to the heliocentric system.}
\tablenotetext{e}{Based on our observations only.}
\tablenotetext{f}{Including additional data available from the literature or by personal communication as follows:
IRAS 07134+1005 -- 
\citet{leb96}, \citet{barth00},
Klochkova (2009 personal communication), \citet{kloch07}, \citet{vanwin00}, \citet{hri03};
17436+5003 --  \citet{bur80};
18095+2704 --  \citet{kloch95};
19475+3119 --  \citet{kloch07}; 
19500$-$1709 -- C. Waelkens (personal communication), \citet{vanwin00};
22223+4327 -- \citet{kloch10}, \citet{vanwin00};
22272+5435 -- R. McClure (personal communication), \citet{zacs09},  \citet{kloch09}, \citet{red02}.
}
\end{deluxetable}

\clearpage

\end{document}